\documentclass[useAMS,usegraphicx,usenatbib]{mn2e}

\usepackage{graphicx}
\usepackage{times}
\usepackage{amssymb}
\usepackage{amsmath}
\usepackage{euscript}

\def\chandra{{\it Chandra}}

\def\s{{\rm\thinspace s}}
\def\ks{{\rm\thinspace ks}}

\def\gyr{{\rm\thinspace Gyr}}
\def\yr{{\rm\thinspace yr}}

\def\Myr{{\rm\thinspace Myr}}

\def\kev{{\rm\thinspace keV}}
\def\erg{{\rm\thinspace erg}}

\def\km{{\rm\thinspace km}}

\def\cm{{\rm\thinspace cm}}

\def\kpc{{\rm\thinspace kpc}}
\def\mpc{{\rm\thinspace Mpc}}

\def\K{{\rm\thinspace K}}
\def\keV{{\rm\thinspace keV}}

\def\msun{{\rm\thinspace M_{\odot}}}

\def\arcmin{{\rm\thinspace arcmin}}

\def\mug{{\rm\thinspace \mu G}}

\def \dd{{\rm d}}

\def\countspacrminsqpkev{{\rm\thinspace counts\thinspace s^{-1} arcmin^{-2} \kev}}
\usepackage{wasysym}

\title[]{The Weak Shock in the Core of the Perseus Cluster}
\author[J. Graham, A.C. Fabian \& J.S. Sanders]{J. Graham$^1$\thanks{E-mail:jgraham@ast.cam.ac.uk}, A.C. Fabian$^1$ and J.S. Sanders$^1$\\
\footnotesize$^1$ Institute of Astronomy, Madingley Road, Cambridge}

\begin{document}

\maketitle

\begin{abstract}
  The dissipation of energy from sound waves and weak shocks is one of
  the most promising mechanisms for coupling AGN activity to the
  surrounding intracluster medium (ICM), and so offsetting cooling in
  cluster cores. We present a detailed analysis of the weak shock
  found in deep \chandra\ observations of the Perseus cluster core. A
  comparison of the spectra either side of the shock front shows that
  they are very similar. By performing a deprojection analysis of a
  sector containing the shock, we produce temperature and density
  profiles across the shock front. These show no evidence for a
  temperature jump coincident with the density jump. To understand
  this result, we model the shock formation using 1D hydrodynamic
  simulations including models with thermal conduction and
  $\gamma<5/3$ gas. These models do not agree well with the data,
  suggesting that further physics is needed to explain the shock
  structure. We suggest that an interaction between the shock and the
  H$\alpha$ filaments could have a significant effect on cooling the
  post-shock gas.

  We also calculate the thermal energy liberated by the weak
  shock. The total energy in the shocked region is about 3.5 times the
  work needed to inflate the bubbles adiabatically, and the power of
  the shock is around $6\times10^{44} \erg \s^{-1}$ per bubble, just
  over $10^{45}\erg \s^{-1}$ in total.
\end{abstract}


\section{Introduction}
The baryonic content of galaxy clusters is dominated by a hot
$10^{7}-10^{8} \K$ plasma---the IntraCluster Medium (ICM)---which
radiates via bremsstrahlung and line emission processes, making
clusters luminous X-ray sources.  Observationally the population of
clusters can naturally be divided into two classes; {\it cool-core}
clusters which show a bright central peak in their surface brightness
and {\it non-cool-core} clusters which do not show sharply peaked
emission. It is believed that the different structure of the two
classes is a result of different merger histories, with cool-core
clusters being more relaxed structures which have developed their
dense cores without significant disruption.

Measurements of the central luminosities of cool-core clusters show
that they have cooling times much shorter than the age of the
universe. This led to the development of the cooling flow model (see
\citealt{Fabian1994} for a review) in which the cooling of gas in
central regions causes it to lose pressure support creating an inflow
of gas toward the centre. The cooled gas was expected to be observed
both as low temperature X-ray emitting gas and as cold gas in the
nucleus of the cluster. However it has long been known that the star
formation rates in central cluster galaxies are below the rate
expected in the cooling flow picture and, with the launch of the
latest generation of X-ray satellites, it has been confirmed that
observed rates of mass deposition from the hot phase are much smaller
than the expected for a cooling flow \citep{Peterson2001,
  Peterson2003}, with the core temperature typically falling to about
a third of the virial temperature. Instead, it is generally accepted
that there is some heat source counteracting the cooling of the X-ray
gas. The nature of this heat source is still an open question but the
Active Galactic Nucleus (AGN) in the cluster centre has emerged as the
leading contender.

Observationally, the centres of cool-core clusters often show highly
disrupted morphologies, in contrast to their overall smooth profile.
Many cool-core clusters shows ``bubbles''---regions with depressed
X-ray surface brightness profiles indicating that they are cavities in the
ICM. These cavities are filled with relativistic plasma and are
evacuated by relativistic jets produced by accretion onto the central
AGN, thereby coupling the gravitational energy released within a few
Schwarzschild radii of the central black hole with the cool-core
region which has a typical diameter of $\sim100\kpc$. Since the rate
of AGN accretion is presumably related to the mass drop out rate from
the ICM, AGN heating has the potential to be self-regulating, thus
explaining the ubiquity of cool-core clusters.

Recent studies of the volumes of observed cavities have found that the
$p\text{d}V$ work done in inflating the cavities is, in most cases,
sufficient to offset the X-ray cooling of the cluster gas
\citep{Birzan2004, Birzan2006, Dunn2006, Rafferty2006}. A natural consequence of
the expansion of these bubbles is the formation of a shock wave which
carries the work done in expanding the bubbles away from the cluster
core. Weak shocks have been observed in several nearby clusters, including
Perseus \citep{Fabian2006} and Virgo \citep{Forman2006}. 

Further evidence of weak shocks is seen in the Perseus cluster as
ripples in the surface brightness \citep{Fabian2003,
  Fabian2006}. These are interpreted to be low-amplitude sound waves
propagating through the cluster. That there are more ripples observed
than bubbles may be an indication that the expansion of the bubble is
not uniform.

The dissipation of weak shock waves at the cluster centre has been
investigated by \citet{Mathews2006} and \citet{Fujita2006}. These authors both
use one-dimensional ideal gas models to conclude that weak shocks
dissipate over a small region compared to the cooling radius of the
cluster and find temperature profiles that are centrally peaked, not
centrally decreasing as observed in real clusters. Nevertheless, the
fact that weak shocks and ripples are indeed observed implies that the
physics of real cluster cores must be more complex than is assumed in
these models.

One of the most surprising conclusions in the deep study of the
Perseus Cluster \citep{Fabian2006} was that the weak shock in Perseus
is isothermal; they found no evidence for the $\sim25$ percent jump in
temperature across the shock expected from the observed density
jump. Moreover, multi-temperature fits showed proportionally more cool
($2\keV$) gas inside the shock than outside, strongly suggesting that
physics beyond that of a simple ideal gas is needed to understand the
structure of the shock and hence energy dissipation in cluster
cores. Here we present a more detailed study of the data first
presented in \citet{Fabian2006}, concentrating on the region around
the inner bubbles in Perseus, including modelling of the shock wave
evolution. We adopt the same cosmological parameters as in that paper;
in particular we assume $H_{0} = 71 \km \s^{-1} \mpc^{-1}$.

\section{Structure of the Shock}

\begin{figure}
  \includegraphics[width=\columnwidth]{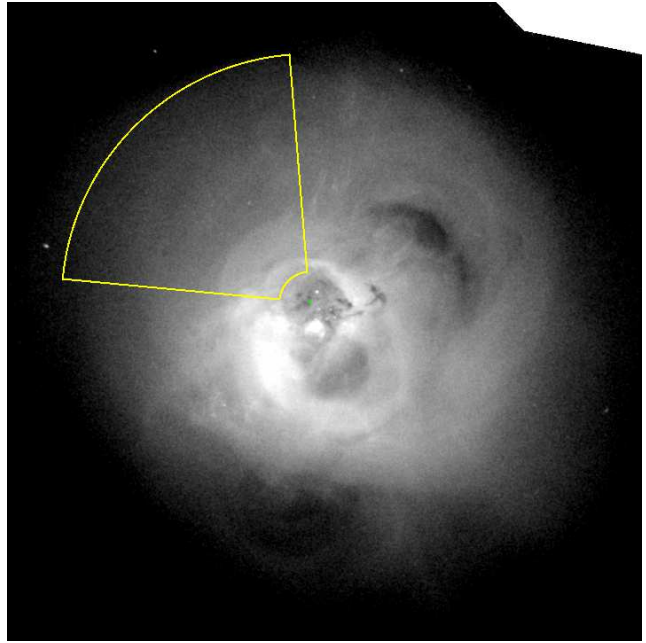}
  \caption{The sector of the Perseus cluster used in the deprojection
    analysis shown on the \chandra\ surface brightness image of the
    cluster core. The sector is chosen to exclude the regions affected
    by cooler projected emission. The centre of the sector is
    positioned so that the shock front is at constant radius, placing
    the centre of the sector at the centre of the bubble.}
  \label{fig:sector}
  \end{figure}

The analysis of the shock front in \citet{Fabian2006} was based on a
projected view of the regions inside and outside of the shock. Whilst
this analysis has the significant advantage that it does not require
any assumptions about the (unknown) geometry of the cluster, the
projected gas can have a significant influence on the results and, in
particular, makes comparison with simulation difficult. To get around
this, we have reanalysed the dataset of \citet{Fabian2006} in the
sector shown in Fig. \ref{fig:sector} performing a deprojection analysis.

\subsection{Surface Brightness Profile}
\begin{figure}
  \includegraphics[width=\columnwidth]{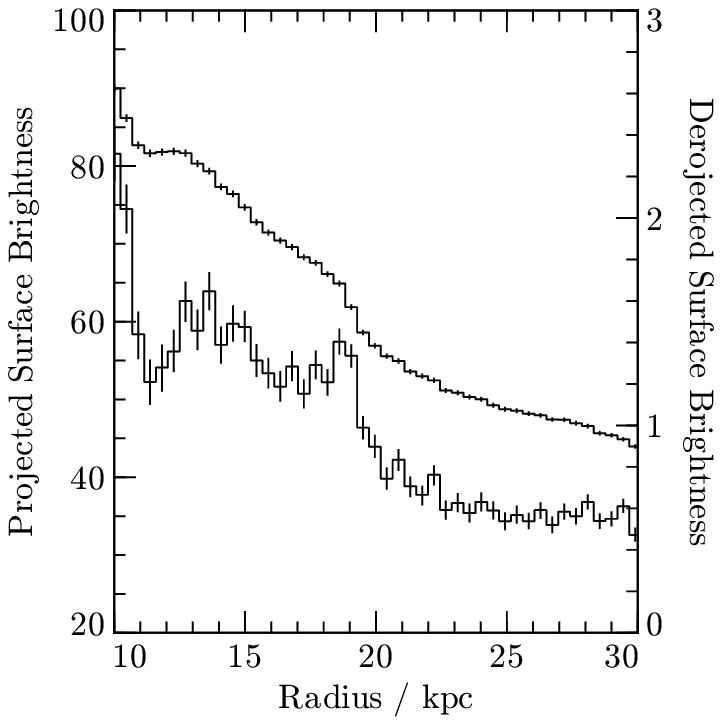}
  \caption{Projected (upper curve) and deprojected (lower curve)
    surface brightness profile in the sector of the Perseus cluster in
    Fig. \ref{fig:sector}. The units are arbitrary.}
  \label{fig:sbshock}
\end{figure}

Fig. \ref{fig:sbshock} shows the projected and deprojected surface
brightness profile in the sector. The sector used to generate the
profile here was quite wide ($80^\circ$). In some places the edge of
the shock appears to be better defined. We examined several sectors
ranging down to $8.9^\circ$, carefully making sure the sector was
aligned along the edge of the shock, but the width of the shock
appeared to remain consistent.

\subsection{Temperature Profile}
  
\begin{figure}
  \includegraphics[width=\columnwidth]{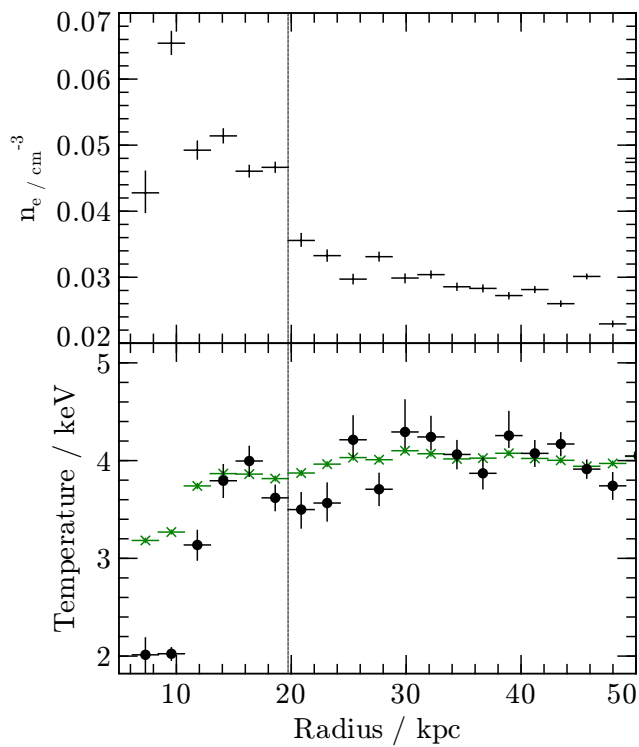}
  \caption{Electron density (top) and projected and
    deprojected (circles) temperature profiles (bottom) in the NE
    sector of Perseus. Radii are measured relative to the centre of
    the sector. The dotted line shows the position of the shock.}
  \label{fig:deprojT}
\end{figure}

Fig. \ref{fig:deprojT} shows the temperature and density profiles
deprojected in $0.1 \arcmin$ ($2.2\kpc$) annuli, using the spectral
deprojection method of \citet{Sanders2007}.  This is a simple method
to calculate deprojected count spectra from observed spectra. Working
from the outside of the cluster, deprojected count spectra are
calculated by subtracting from the observed projected spectrum the
contribution from those shells which lie between the emitting region
and the observer. Further details of the deprojection method,
including a comparison with other deprojection codes such as
\textsc{ProjCT} will be provided in a future paper (Sanders, et al.,
in prep.). For our analysis the spectra have been grouped to a minimum
of 100 counts per spectral bin in the outside annulus, with the others
having the same binning. Spectral fits are performed by minimising the
$\chi^2$ statistic, with the temperature, column density, abundance
and normalisation held free, and the redshift frozen at $0.0183$.

At the shock front, the density jumps by a factor $1.31\pm0.04$. The Mach
number and expected temperature jump can then be derived from the
Rankine-Hugoniot relations, assuming the curvature of the shock front
has a negligible effect. With $\gamma=5/3$, appropriate if the ICM is
gas (as opposed to cosmic ray) dominated, the Mach number of the shock
is $1.21\pm0.03$ and the expected temperature jump a factor
$1.27\pm0.03$.

In the deprojected temperature profile, it is apparent that there is
no significant temperature jump between the first preshock annulus and
the first postshock annulus. The temperature ratio is
$1.03\pm0.06$. This is clearly inconsistent with the prediction of a
purely hydrodynamic model.

Although there is no temperature jump at the shock, there is some
evidence for a temperature peak $\sim3\kpc$ behind the shock
front. The ratio of this maximum in the postshock temperature and the
minimum of the preshock temperature is $1.14\pm0.07$, which is
marginally consistent with either a flat profile or a standard shock
jump. However, interpreting this feature as a standard shock requires
some process beyond simple hydrodynamics to explain the offset between
the shock face and the temperature jump.

The annulus centred around $11.8\kpc$, shows a drop in both the
temperature and the density, corresponding to a drop in the gas
pressure. The origin of such a drop is not well understood; it
suggests an additional non-thermal contribution to the pressure from
e.g. magnetic fields, cosmic rays, or ram pressure in this
region. Taking only points outside this, the overall temperature
profile shows evidence for some structure with a constant model
providing a poor fit (reduced $\chi^2=2.7$).

Accounting for the need to fit both the actual structure of the
shock front, and the fact that the temperature profile far in front
and behind the shock is not well modelled, we choose to proceed by
comparing models profiles to the temperature points in the region,
$16-25\kpc$. This encompasses two annuli inside and three annuli
outside the shock.

\begin{figure}
  \includegraphics[width=\columnwidth]{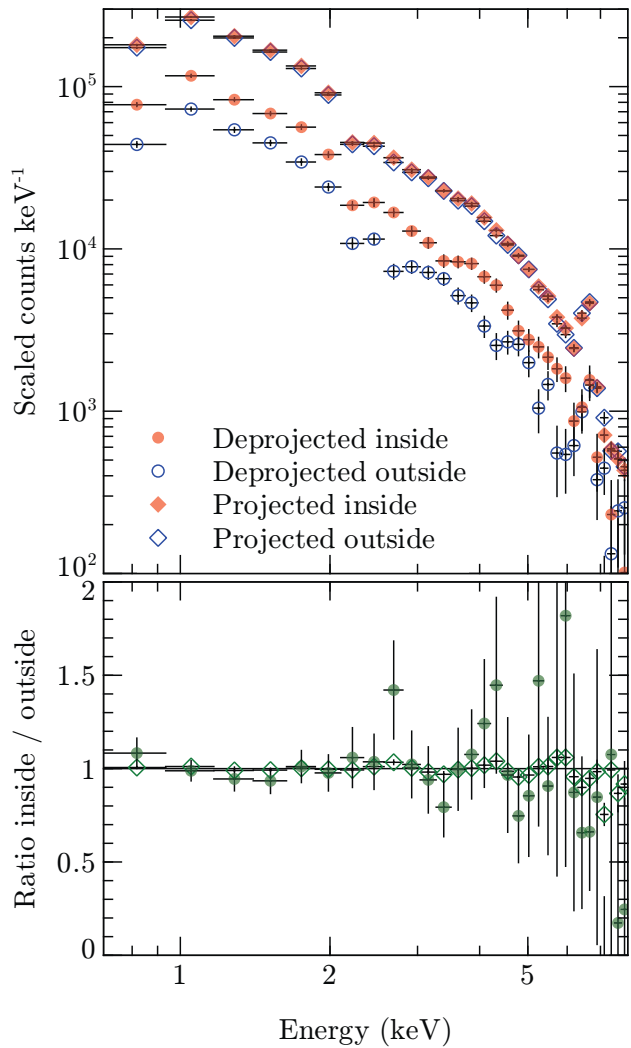}
  \caption{Comparison of the projected (diamonds)and deprojected
    (circles) spectra in the immediately inside (filled red points)and
    immediately outside the shock (unfilled blue points). The
    projected spectra are presented without any scaling, whilst the
    deprojected counts (circles) are scaled to units of $10^{-5}
    \countspacrminsqpkev$, to take into account the area of the
    sector. The ratio plot (bottom) gives the ratio of the counts per
    unit energy inside the shock to those outside, scaled to have the
    same overall normalisation. The projected spectra inside and
    outside the shock show remarkable consistency.}
  \label{fig:spectra_ratio}
\end{figure}

\subsubsection{Checks on the Deprojection Analysis}

In order to check that the low apparent temperature jump behind the
shock is not an artifact of the deprojection procedure, we have
compared the spectra from the annuli either side of the
shock front.  Fig. \ref{fig:spectra_ratio} shows the projected and
deprojected spectra from regions immediately inside and outside the
shock front. The spectra on each side of the shock are remarkably
similar, both in projection and in deprojection, consistent with the
small temperature jump found from the full deprojection analysis.

An additional test we made was to use the spectrum from outside the
shock as a background for the shocked region. This is almost certainly
an overestimate for the amount of emission projected onto the shocked
region. Using this background, the measured temperature was still
compatible with our deprojected results. It appears not to matter how
much projected emission is subtracted from the shocked region as the
spectra are almost identical.

\subsubsection{Non-Equilibrium Ionisation}

\begin{figure}
  \includegraphics[width=\columnwidth]{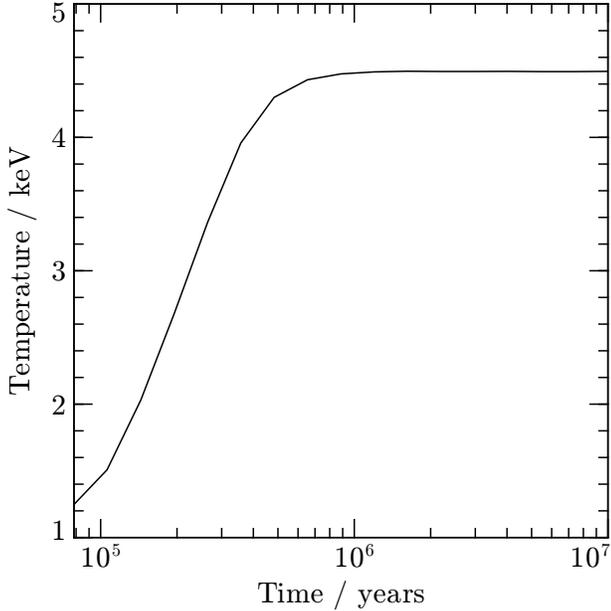}
  \caption{Fitted temperature for {\sc nei} models with $n_{e}=0.03$ and
    $T=4.5\kev$ at various times after the shock has passed.}
  \label{fig:nei}
\end{figure}

The models fitted above assume that the ions in each zone are in their
equlibrium ionisation state. Following the rapid temperature change
induced by the passage of a shock, there is a finite time before
ionisation equlibrium is re-established.  During this time, the
temperature inferred from X-ray spectral fitting will be lower than
the actual gas temperature, suggesting a possible explanation for the
offset of the temperature jump at the shock front.  This was
tested using the {\sc xspec} {\sc nei} model \citep{Hamilton1983,
  Borkowski1994, Liedahl1995, Borkowski2001} with a {\sc phabs}
absorbed component to simulate \chandra\ observations at various times
after the shock passage and fitting the simulated data with the {\sc
  equil} model. The temperature fitted to these models is shown in
Fig. \ref{fig:nei}; clearly at $t=10^{6}\yr$ the fitted temperature
is equal to the input temperature. Since the shock velocity is
approximately $1\kpc/10^{6}\yr$ non-equlibrium ionisation effects can
account for an offset of no-more than $1\kpc$, indicating that
non-ionisation equlibrium alone is not enough to explain the apparent
density/temperature offset at the shock front.

\section{Modelling AGN-induced Weak Shocks}

\subsection{Equations of fluid dynamics}

For the purposes of our model, we assumed the dynamics of the ICM
are those of a single phase, inviscid fluid in which magnetic fields
are too weak to be dynamically important and where the electron-ion
coupling time is much shorter than any timescale of interest. In this
case the equations describing the evolution of the ICM are those of
ideal hydrodynamics (e.g. \citealt{LandauLifshitzFluids}):

\begin{align}
\frac{D \rho}{D t} + \rho \nabla \cdot u = 0
\label{eqn:fluids_mass}\\
\rho \frac{D u}{D t} = -\nabla p - \rho g
\label{eqn:fluids_momentum}\\
\frac{D \epsilon}{D t} + p\frac{D
  \left(\frac{1}{\rho}\right)}{Dt} = -\frac{1}{\rho}\nabla \cdot \left({\kappa \nabla T}\right)
\label{eqn:fluids_energy}
\end{align}
where $D \equiv \frac{\partial}{\partial t} + u \cdot \nabla$ is the
total (Lagrangian) derivative, $u$ is the fluid velocity, $p$ the
pressure, $\rho$ the density, $\epsilon$ the specific internal energy,
$T$ the temperature, and $\kappa$ the conductivity. For a fully
ionised hydrogen plasma, the conductivity $\kappa$ is given by
\citep{SpitzerFullyIonized}:

\begin{equation}
\kappa =
640\epsilon\delta_\text{T}\left(\frac{2\pi}{m_\text{e}}\right)^{1/2}\frac{k_\text{B}\epsilon_0^2}{e^4}
\frac{(k_\text{B}T)^{5/2}}{Z\ln\Lambda}
\label{eqn:spitzer_conductivity}
\end{equation}where Z is the charge of the ion, $\Lambda$ is the Coulomb logarithm
with $\ln \Lambda \approx 37$ in the conditions of interest,
$\epsilon\approx0.4$ accounts for the reduction of conductivity in a
plasma due to the electric field and $\delta_\text{T}\approx0.225$
accounts for the finite proton mass. The remaining symbols have their
usual physical meanings.

In the case of a steep temperature gradient and strong conductivity,
the rate of heat flow predicted from the bulk description above can
exceed the maximum rate at which the electrons are able to transport
energy. In this case it is appropriate to replace the conduction term
in the energy equation \eqref{eqn:fluids_energy} by one appropriate for saturated conductivity \citep{CowieMckee1977}:

\begin{equation}
\frac{D \epsilon_\text{conduction}}{D t} \approx - \frac{1}{\rho}\nabla
\cdot \left( 0.4 \left( \frac{2}{\pi m_{e}} \right)^{1/2} n_{e} (k_{B}T)^{3/2}\right)
\end{equation}

\subsection{Shock Propagation in the ICM}

In general, propagation of a shock wave in a highly ionised plasma such
as the ICM differs from that in an ordinary fluid because of the very
different sound speeds for the electrons and ions. At the shock front,
the ions undergo a temperature and density jump given by the normal
Rankine-Hugoniot relations for a shock of Mach number $M$:

\begin{align}
\frac{\rho_{2, \text{shock}}}{\rho_{1,\text{shock}}} = \frac{\gamma +
  1}{(\gamma-1)+\frac{2}{M^2}}
\label{eqn:shock_density}
\\
\frac{T_{2, \text{shock}}}{T_{1,\text{shock}}}=
\frac{\left( 2\gamma M^2 - (\gamma-1) \right) \left( 2+(\gamma-1)M^2\right)}{(\gamma+1)^2 M^2}
\label{eqn:shock_temperature}
\end{align}

Charge neutrality requires that the
electrons are compressed adiabatically at the shock front in the same
ratio as the ions, leading to a temperature difference:
\begin{equation}
T_\text{i} - T_\text{e} = \frac{T_{2, \text{shock}}}{T_{1,
    \text{shock}}} - \left( \frac{\rho_{2, \text{shock}}}{\rho_{1,
    \text{shock}}} \right)^{\gamma-1}
\label{eqn:temp_diff}
\end{equation}

The two components are brought into thermal equilibrium behind the
shock over a timescale comparable to their thermal equilibration time
\citep{ZeldovichRaizer}:
\begin{equation}
  \tau_\text{ei} \approx = \frac{1.1\times10^{16} M A}{(n_i/\cm^{-3}) Z^2
    \ln \Lambda}
  \left(\frac{T}{\kev}\right )^{3/2}
\end{equation}

Where $\tau_\text{ei}$ is the
electron-ion equilibration timescale, $A$ the atomic mass of the ions,
and $n_i$ the number density of ions. For the case of the Perseus cluster
core with $n_i\sim0.03\cm^{-3}$, $T\sim3.5\kev$, $\bar{A}=1.29$,
$\bar{Z^2}\approx\bar{Z}^2=1.2$ and $M=1.2$, electron-ion
equilibration occurs approximately $3.2\kpc$ behind the shock,
consistent with the apparent offset between the temperature and
density jumps at the shock front. In the case of a weak shock,
however, the heating is largely due to adiabatic compression which acts
equally on the electron and ion components and so from
\eqref{eqn:temp_diff} the temperature difference is only
$\sim1$ percent. Therefore electron-ion non-equilibration is unlikely to
explain the observed density/temperature offset and we can approximate the
behaviour of the ICM in the shocked region as that of a
single-component fluid.

\subsection{Conductivity of the ICM}

In general the conductivity of the ICM may differ from the Spitzer
value of equation \eqref{eqn:spitzer_conductivity} because even
dynamically unimportant magnetic fields are effective in suppressing
transport across field lines. Therefore the level of conductivity is
strongly affected by the geometry of the field. In general it is
assumed that the field is sufficiently tangled that, on scales of
interest, isotropic conduction occurs due to motion of electrons along
diverging field lines and so the conductivity is the Spitzer form
reduced by some factor $f<1$; \cite{Narayan2001} derive a suppression factor of
$f=0.2$ based on the assumption of a tangled field. If the field is
more ordered on smaller scales, the conduction will not be isotropic
and parallel to the field lines the conductivity will be close to the
full Spitzer value.

In the case of the ICM, there is some evidence of ordered field
structures on scales $>$ a few $\kpc$. Surrounding many central
cluster galaxies are bright, line emitting nebulae. In several cases
including the Perseus \citep{Conselice2001} and Centaurus clusters,
long, largely radial, filamentary structures (up to $60\kpc$ in
projection), are observed which are apparently long lived despite
being embedded in the hot ICM \citep{FabianHalpha, Hatch2005}.  In
order to be stabilised against evaporation by the surrounding gas,
thermal conduction between the filaments and their surroundings must
be suppressed by about an order of magnitude more than in a tangled
field \citep{NipotiBinney2004}. This is suggestive of a field geometry
in which the filaments are enclosed in a sheath of radially-coherent field
lines.

Increasing radial coherence of the magnetic field toward the centre of
the cluster is expected if there is radial flow of gas with a
radially decreasing velocity amplitude, for
example in a (reduced) cooling flow
\citep{SokerSarazin1990}. Recent numerical studies of bubbles evolution
in a magnetised ICM has also shown the magnetic field becomes more
ordered in the wake of the rising bubble \citep{Ruszkowski2007}. With
these results in mind, we assume that conduction may proceed radially at up to
the full Spitzer rate.

\subsection{Numerical Method}

We have written the fluid equations \eqref{eqn:fluids_mass} --
\eqref{eqn:fluids_energy} in a form appropriate to a one dimensional,
spherically symmetric system:
\begin{align}
\frac{\dd r}{\dd t} = u\\
\frac{\dd u}{\dd t} = -\left(g+\frac{1}{\rho}\frac{\dd p}{\dd r}\right)\\
\frac{\dd \epsilon}{\dd t} = -p\frac{\dd}{\dd
  r}\left(\frac{1}{\rho}\right) -\frac{1}{\rho r^2}\frac{\dd}{\dd
  t}\left(r^2 q \right)
\end{align}
where $q$ is the appropriate conductive heat flux. These are
supplemented with the ideal gas equation of state:
\begin{equation}
p = \frac{\rho}{\mu m_H} k_B T
\end{equation}
where we have taken the average particle mass $\mu$ as a constant
set to $0.61$ appropriate to the chemical composition of the ICM.

To numerically solve these equations, we used an explicit finite
difference method on a Lagrangian grid, based on the method of
\citet{RichtmyerMorton} and following the implementation of
\citet{Morris2003}. An artificial viscosity term was added to the
equation of motion to correctly model the shock jump conditions. The
strength of the artificial viscosity was chosen so that the shock was
smeared over approximately four grid zones.

For simplicity we neglect the contribution of the cluster galaxies to
the gravitational acceleration and assume the contribution from gas is
a fixed fraction of the dark matter contribution. Assuming the dark
matter follows a NFW profile \citep{NFW1997}, the acceleration may be
written:

\begin{equation}
g_\text{NFW} =
\frac{GM_\text{virial}}{r^2} \frac{\log(1+cr/r_\text{virial}) - 
cr/(r_\text{virial}+cr)}{\log(1+c) - c/(1+c)},
\end{equation}
where $M_{\text{virial}}$ and $r_{\text{virial}}$ are the virial mass and
radius respectively and $c$ is the cluster concentration.

The shock wave was generated by moving the position of the inner
boundary. If the jet supplies energy at a constant rate $\dot{E}$ then
it can be shown that the bubble radius varies with time as:

\begin{equation}
R_{b} = \left (\frac{\dot{E}t^3}{\rho} \right)^{\frac{1}{5}}.
\end{equation}

Numerically this formulation is problematic as it requires a high
(formally infinite) velocity at $t=0$. Instead we have adopted a
simpler model with an initial exponential increase in the expansion
velocity to a maximum and a subsequent fall off like $\cos (t)$:

\begin{equation}
u_\text{piston} = 
\begin{cases}
u_0\left(\frac{\exp\left (t/t_\text{smooth}\right )-1}{e-1}\right) & {0 < t <
    t_\text{smooth}}\\
u_0\cos\left(\frac{2\pi}{t_p}\left(t-t_\text{smooth}\right ) \right ) &
{t_\text{smooth} < t < t_\text{max}}\\
0 & t > t_\text{max}
\end{cases}
\end{equation}

The smoothing time $t_\text{smooth}$ has been fixed at $1\Myr$; with
this time, no spurious central heating due to sharp changes in velocity are
observed. $t_p$ is a characteristic piston timescale and $u_0$ defines the
maximum piston velocity. Since we are concerned with only a single
injection event, we choose $t_\text{max}$ to occur when the piston
velocity first reaches zero, i.e. $t_\text{max}=t_{\text{smooth}} + t_p/4$.

\section{Comparison with Data}

\subsection{Initial Conditions}

\begin{figure}
  \includegraphics[width=\columnwidth]{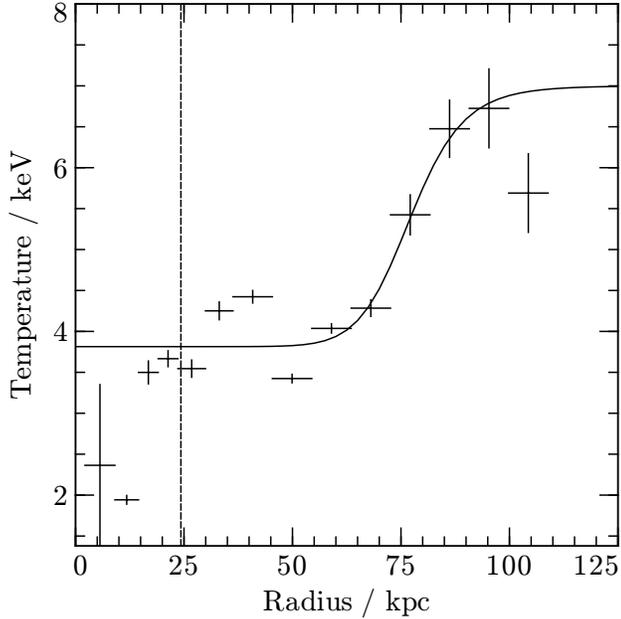}
  \caption{Deprojected temperature data from the 200ks observation
    \citep{Sanders2004} for a sector of the Perseus cluster containing
    the shock (points) and adopted initial temperature profile
    (line). The sector used contains the four regions covering the
    shock in Fig. 18 of \citep{Sanders2004}. This region is similar to
    that in Fig. \ref{fig:sector} but has a significantly greater radial
    extent. The vertical line shows the approximate position of
    the shock.}
  \label{fig:temperature_perseus}
\end{figure}

As we are primarily concerned with modelling the observed shock in the
Perseus cluster, the majority of our models are set up with properties
close to those of the NE sector of the Perseus cluster used in the
deprojection analysis. Our initial temperature profile was based on
the observed temperature in the pre-shock region (i.e. $r>20\kpc$). To
determine the shape of the profile at karge radii, we used a fit to
the $200\ks$ data presented in \citet{Sanders2004} covering the shock
region, whilst the temperature in the immediate vicinity of the shock
was matched to the data in Fig. \ref{fig:deprojT}. The resulting
profile, together with the \citet{Sanders2004} data is shown in
Fig. \ref{fig:temperature_perseus}. The functional form is:

\begin{equation}
T = 7\frac{1+\left(\frac{r}{73\kpc}\right)^{14}}
{1.93+\left(\frac{r}{73\kpc}\right)^{14}}\kev
\end{equation}

The gravitational acceleration was modelled using the NFW component of
the profile in \cite{Mathews2006}, which provides an acceptable fit to
the observed acceleration in the region $r>10\kpc$ which is of
interest here. The initial density profile is calculated assuming
hydrostatic equilibrium and a central density $n_e = 0.05
\cm^{-3}$. This gives a gas mass fraction inside $25\kpc$ of $\sim6\%$,
so a static gravitational field is expected to be a good
approximation. The piston timescale $t_p$ was taken as $1\times 10
^{7}\yr$ and the smoothing time as $1\times 10^6 \yr$.

The inner boundary was started at $r=5\kpc$ --- approximately the
distance from the bubble centre to the cluster centre --- and the piston
amplitude adjusted to $r_\text{piston}=3\kpc$ so that the shock
approximately reproduced the expected density jump at the current
shock position.

When adding conduction to the models, Spitzer fractions, $f$, in the
range 0-2 have been considered. Cases with the conductivity in excess
of the Spitzer value are intended as a simple approximation for a
situation where the electrons and ions are decoupled. In this case
only $\sim1/2$ of the energy needs to be transported to decrease the
observed (electron) temperature compared with the fully coupled case.

\subsection{Simulation Results}

\begin{figure}
  \includegraphics[width=\columnwidth]{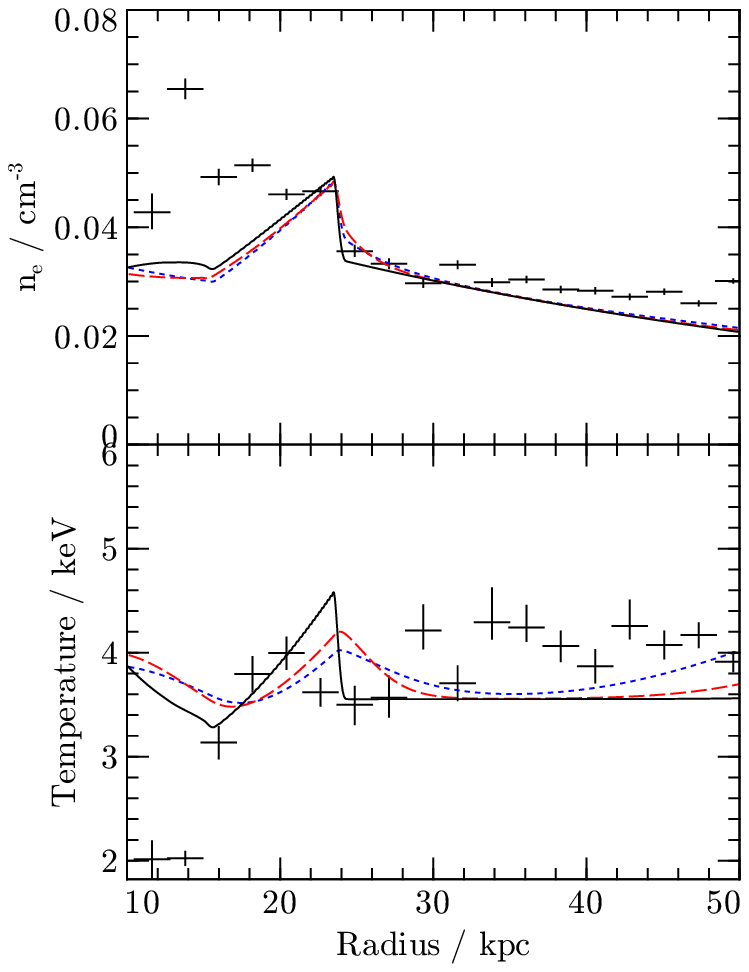}
  \caption{Density and temperature after the shock has reached $24\kpc$
    with $\gamma=\frac{5}{3}$ and no conduction (solid line), a Spitzer
    fraction of 1.0 (dashed line) and 2.0 (dotted lines) and
    deprojected observations in the NE sector (points).}
  \label{fig:results_perseus_T}
\end{figure}

For the purposes of comparison with the simulations, we have used
radii measured from the cluster centre rather than from the centre of
the deprojected sector, since this more closely matches the setup of
the simulated cluster. The radius of the shock is then $24\kpc$.

Fig. \ref{fig:results_perseus_T} shows the density and temperature
structure of a simulation with $\gamma=5/3$ after the shock has
reached the observed radius (taken to be $24\kpc$), The time taken for
this to occur was dependent on the level of conduction; in the non
conductive case the time was $\sim 12\Myr$, stronger conduction
reduces the pressure difference across the shock so slowing its
propagation. The piston amplitude was adjusted so that the observed
density jump was matched in each case.  The addition of conduction
with $f=1$ and $f=2$ reduces the height of the jump to about 65 percent and
50 percent respectively of the non-conductive case. However conduction does
not appear to change the position of the temperature jump relative to
the density jump.

A more quantitative comparison of the data and the models can be made
by performing a $\chi^2$ fit of the model temperature profile to
the data in the region surrounding the shock. To allow for the fact
that the data shows more structure ahead of the shock than the initial
conditions of our model, we have selected only those data points in the
region $18\kpc \leq r \leq 28\kpc$ and spectrally averaged the model
temperatures to match the data bins. A simple fit of the models to the
data products a rather poor fit in each case with the reduced $\chi^2$
ranging from 4.1 in the case $f=2$ to 7.2 in the case $f=0$. The
quality of the fit also depend somewhat on the initial temperature profile
assumed. For example, maintaining the shape of the profile but
increasing the central value makes the fits poorer in all cases.


We have also considered models in which the ratio of specific heats is
less than the standard ideal gas value of $5/3$. Lower values of $\gamma$ may be
appropriate if the ICM in the neighbourhood of the shock is
significantly contaminated by relativistic particles diffusing from
the bubbles \citep{Mathews2007, Ruszkowski2007a}. In this case the
effective adiabatic index is given by:
\begin{equation}
\gamma_\text{eff} = \frac{\gamma_\text{Th} + X_\text{CR}\gamma_\text{CR}}{1+X_\text{CR}}
\end{equation}
with $X_\text{CR} = p_\text{CR}/p_\text{Th}$ \citep{Pfrommer2007}. We consider
$\gamma=4/3$ as a limiting case of cosmic-ray-pressure dominated gas
and $\gamma=3/2$, i.e. $X_\text{CR}=1$, as a realistic value for the
core of the Perseus cluster based on the analysis of the non-thermal
emission by \citet{Sanders2007}.

\begin{figure}
  \includegraphics[width=\columnwidth]{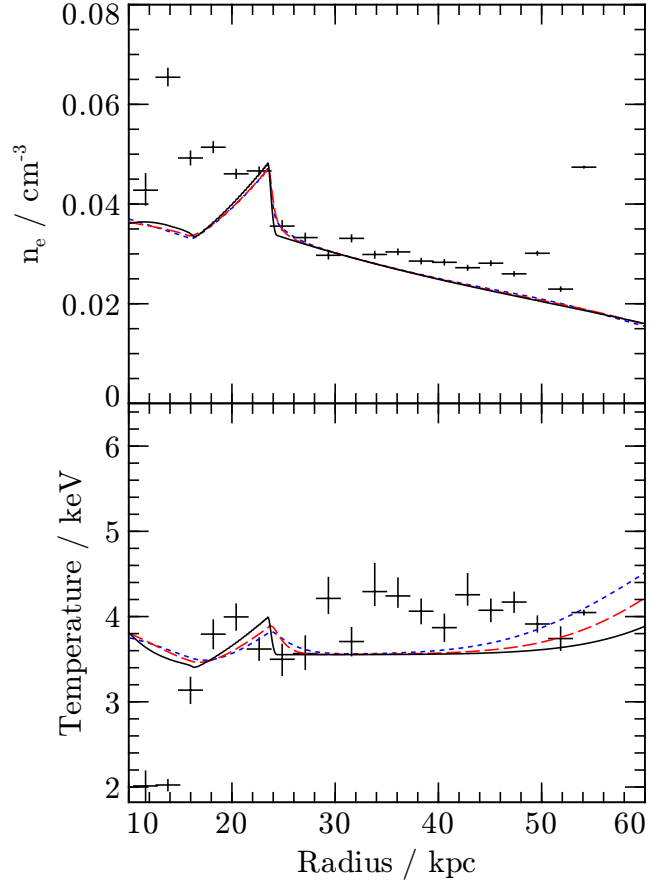}
  \caption{Shock model at $r=\kpc$ for the Perseus-like model cluster
    with $\gamma=\frac{4}{3}$ and no conduction (solid line), a Spitzer
    fraction of 1.0 (dashed line) and 2.0 (dash-dotted line).}
  \label{fig:results_perseus_lowgamma}
\end{figure}

The temperature profiles for the $\gamma=4/3$ case are shown in
Fig. \ref{fig:results_perseus_lowgamma}. As expected from the
standard shock jump conditions, the magnitude of the shock in the is
substantially decreased compared to the $\gamma=5/3$ case. However
conduction appears to show little further effect in this
case. The reduced $\gamma$ appears to somewhat improve the agreement
between simulation and data but does not produce a statistically
acceptable fit in either case.

\begin{table}
\begin{tabular}{|c|c|c|c|c|}
\hline
\multicolumn{4}{c}{Model Parameters} & \multicolumn{1}{c}{Fitting results}\\
$\gamma$ & $r_\text{p}/\kpc$ & $t_\text{p}/10^{7}\yr$ & $f$ 
& Reduced $\chi^2$ \\
\hline
$5/3$ & 2.75    & 1   & 0 & 7.4  \\
$5/3$ & 3       & 1   & 1 & 5.6  \\
$5/3$ & 3       & 1   & 2 & 4.1  \\
$3/2$ & 2.5     & 1   & 0 & 3.7  \\
$3/2$ & 2.5     & 1   & 1 & 3.4  \\
$3/2$ & 2.5     & 1   & 2 & 3.1  \\
$4/3$ & 2.5     & 1   & 0 & 2.3  \\
$4/3$ & 2.6     & 1   & 1 & 2.6  \\
$4/3$ & 2.6     & 1   & 2 & 2.4  \\
$5/3$ & 2.75, 2 & 1,1 & 0 & 10.7 \\
$5/3$ & 3, 2    & 1,1 & 1 & 7.7  \\
$5/3$ & 3, 2    & 1,1 & 2 & 5.7  \\
\hline
\end{tabular}
\caption{Results of numerical simulations for various values of
  $\gamma$, the Spitzer fraction $f$ and numbers of injection
  events. The piston amplitude $r_\text{p}$ is adjusted so that the
  density jump at $24\kpc$ matches the observations. The quoted
  reduced chi-squared is based on a fit of the temperature data alone
  to the five data points in the region $18-28\kpc$ enclosing the
  shock front.}
\label{table:model_results}
\end{table}

\subsubsection{Models with Multiple Consecutive Outbursts}\label{sec:multiInject}

By construction our models match the density and the density jump at
the shock front. Nevertheless, the models considered so far all show
high density regions inside the shock considerably narrower than that
observed. 

\begin{figure}
  \includegraphics[width=\columnwidth]{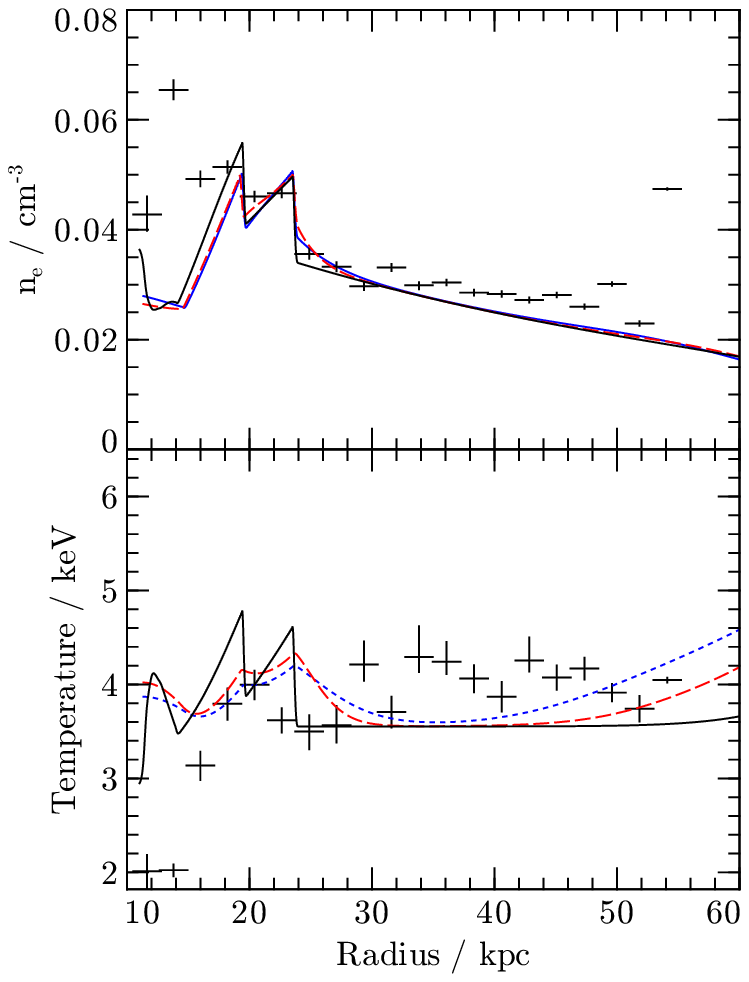}
  \caption{Temperature profile for a model with two episodes of
    injection starting at $t=0$ and $4.5\times 10^{6}$ with $r_p=3\kpc$
    and $2\kpc$ respectively and $t_p=1\times10^{7}\yr$ in each case.}
  \label{fig:results_perseus_multiinject}
\end{figure}

One way of addressing this issue is to invoke multiple outbursts
during a single cycle of AGN activity. Fig.
\ref{fig:results_perseus_multiinject} shows the density and
temperature profiles for set of models in which the single outburst
has been replaced by two identical outbursts beginning at $t=0$ and
$t=4.5\Myr$ .

This model successfully reproduces a thicker rim of high density gas
that extends in approximately as far as the obscuring cool gas seen in
the data, although it tends to produce a somewhat worse fit to the
temperature profile than the single-outburst models.

\subsection{The Temperature Offset}

Since our simulations do not convincingly reproduce a
temperature/density offset at the shock front, we must consider other
mechanisms for cooling the gas in the immediate aftermath of the
shock. 

\subsubsection{Non-Uniform Initial Temperature}

\begin{figure}
  \includegraphics[width=\columnwidth]{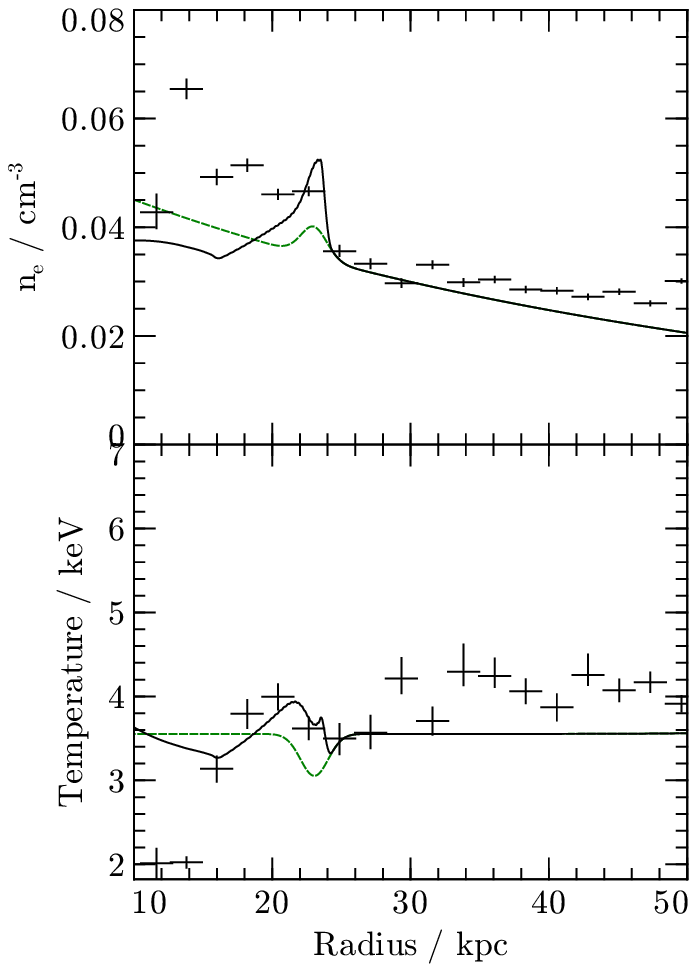}
  \caption{Density (top) and temperature (bottom) profiles at $t=0$
    (dashed lines) and $t=13\Myr$ (solid lines) for a model with a
    blob of colder gas close to the observed shock radius.}
  \label{fig:results_perseus_dip}
\end{figure}

We have assumed a temperature profile that is flat in the core of the
cluster. However the deprojected temperature profile shows evidence
for oscillations in the temperature of up to $\sim 0.5\keV$. A blob of
cold gas producing a temperature fluctuation of around this magnitude
at the position of the observed density jump can produce density and
temperature profiles similar to the observed ones--- a simulation of
such a profile is seen in Fig.
\ref{fig:results_perseus_dip}. However this relies on fine-tuning of
the initial temperature profile, suggesting that weak shocks in
general would not show such an offset. However, the very high
resolution observations required to test this are only likely to be
possible for a very small number of local clusters in the foreseeable
future.

\subsubsection{Mixing of Filaments}

\begin{figure*}
  \includegraphics[width=\textwidth]{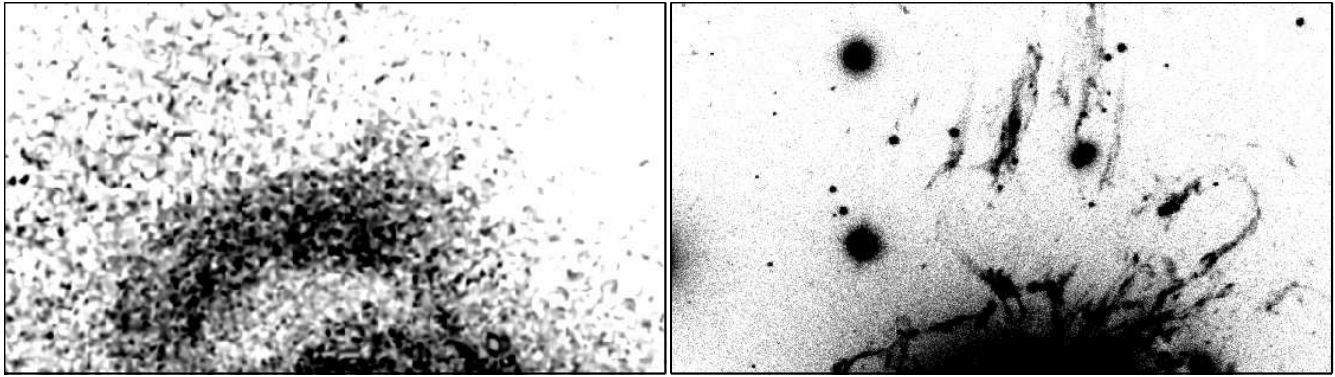}
  \caption{X-ray pressure (\citealt{Fabian2006}, left) and H$\alpha$
    (\citealt{Conselice2001}, right) maps for a 150$\times$ 80
    arcsecond region in the north of of the Perseus cluster. The
    filaments in the H$\alpha$ map appear to terminate at the shock
    front in this region}
  \label{fig:filaments}
\end{figure*}

Thus far, we have ignored the multiphase nature of the Perseus core in
our analysis. Of particular significance is the reservoir of cool gas
contained in the large filamentary nebula associated with the central
cluster galaxy. However, there is observational evidence for an
interaction between the phases; Fig. \ref{fig:filaments} shows the
X-ray pressure and $H\alpha$ map around the region we have
analysed. The filaments in this region appear to terminate at the
shock front, suggesting the cool gas from the filaments is being mixed
in with the hot ICM in this region. This mixing could have a
significant impact on the temperature structure of the shock.

The filaments are known to contain ${\rm H}_2$ with a temperature of
$300-10000\K$ and a mass per unit length of around $10^{4} - 10^{5}
\msun/\kpc$ \citep{Johnstone2007}. Associated with the filaments is
soft X-ray emission \citep{FabianHalpha}, tracing a total mass of
$\sim10^9 \msun$ within the inner $1.5\arcmin$ ($\sim33 \kpc$)
\cite{Fabian2006}. There is also a detection of CO in the region of
the filaments \citep{Salome2006}, implying a population of gas at
10-100K, although it is not clear if this is confined to the filaments
or more dispersed. If it is indeed confined to the filaments it
implies a mass per unit length of around $10^{8}\msun/\kpc$. The
density of the X-ray emitting gas in the region of the shock is
approximately $7\times10^5 \msun/\kpc^3$ so, assuming complete mixing
of the behind the shock, a single filament could cool around
$1000\kpc^2$ of shock front by $0.5\keV$, an area comparable to the
surface area of a spherical shock centred on the Perseus bubbles and
at the observed radius. However, in practice the efficiency of mixing
is likely to be considerably less than 100 percent, so this estimate is an
upper bound.

\begin{figure}
  \includegraphics[width=\columnwidth]{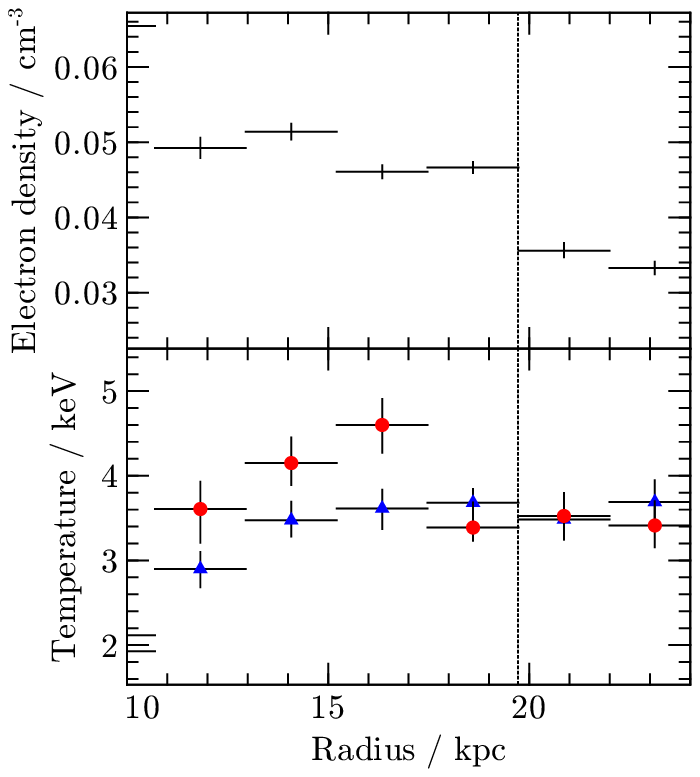}
  \caption{Full sector density profile (top) and split sector
    temperature profiles (bottom). The vertical line marks the
    position of the shock. The circular points are for the NW most
    half of the original sector and the triangular points for the
    SE. A jump in temperature behind the shock front is seen only for
    the NW-most sector.}
  \label{fig:two_sector}
\end{figure}

If the post-shock gas is cooled by mixing of filaments, then we might
expect a spatial variation in the post-shock temperature according to
the distribution of filaments along the shock
front. Fig. \ref{fig:two_sector} shows the deprojected density and
temperature profiles when the sector in Fig. \ref{fig:sector} is
divided into two equal-angle sectors. Outside the shock front the two
profiles are consistent, but inside the shock there is a significant
difference in the temperatures for the two sectors; one is consistent
with a flat temperature profile whilst the other has a jump of
$1.36\pm0.13$ between the preshock minima and postshock
maxima. Interestingly, it is the sector closer to the currently
observed filaments that has the larger temperature jump; this may be
indicative of the shock having destroyed previous filaments in the
cooler sector.

Assuming the filaments have had a significant effect on the
temperature structure of the gas, it is not clear why the postshock
temperature should be roughly the same as the preshock temperature; by
mixing more or less cold gas into the ICM this  could be varied from
the postshock temperature expected from the Rankine-Hugoniot relations
to a temperature significantly lower than the preshock temperature. A
physical mechanism to explain the near-isothermality observed would
make this model significantly more appealing.

\subsection{Repeated Shock Heating}

\begin{figure}
  \includegraphics[width=\columnwidth]{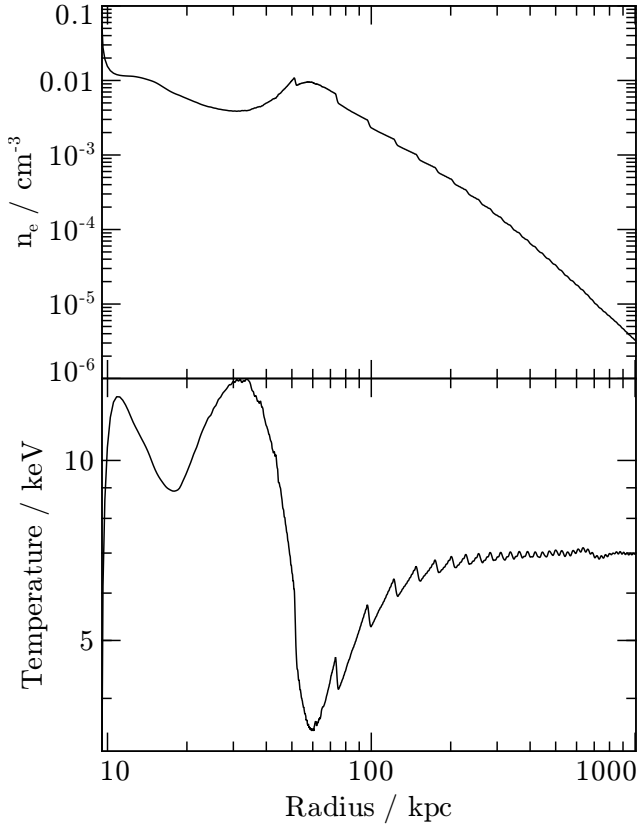}
  \caption{Density and temperature profiles for a simulation in which
    a Perseus-like cluster is subject to monochromatic wave heating
    for $4\gyr$. The central temperature peak is characteristic of
    this kind of simulation but is inconsistent with the data.}
  \label{fig:long_run}
\end{figure}

We now briefly address the issue of repeated shock heating of the
cluster core. Previous authors \citep{Mathews2006, Fujita2006}, have
found that repeated shocking of the cluster core leads to a centrally
increasing temperature profile, clearly inconsistnet with the
observations. This is a result that we can reproduce in our model;
Fig.  \ref{fig:long_run} shows the result of repeatedly shocking a
cluster, with the initial conditions identical to those employed for
the models of the Perseus shock, over a period of $4\gyr$ with wave
parameters $r_p=2.75\kpc$, $t_p=2\times10^{7}\yr$ and
$r_\text{min}=10\kpc$.  Clearly the temperature profile produced is
not the smoothly decreasing temperature profile we expect from
observations.

However, we recommend caution in a literal interpretation of these
results derived from these 1D models. There are several limitations of
the models that may cause the discrepancy between the model and the
data:

\begin{itemize}
\item The 1D nature of the models misses several essential features of
  real clusters. The geometry of bubble heating in 3D allows the
  redistribution of material from outside the bubble into the core of
  the cluster and conversely the entrainment of the material behind
  the rising bubble draws material away from the centre
  \citep[e.g.][]{Gardini2007}. Changes in the jet direction and bulk
  motion of the ICM in the cluster potential are also significant
  effects in ensuring that the same cluster gas is not repeatedly
  subject to the strongest heating, as happens in the 1D case
  \citep{Bruggen2005, Heinz2006}.

\item The assumption of constant piston amplitude (``bubble size'')
  over time is likely to be unrealistic. One would expect a real
  source to release bubbles with a variety of sizes over a variety of
  timescales. Evidence of this in the Perseus cluster is seen in the
  spatial variation of the ripple amplitude reported by
  \citet{Sanders2007}. In addition to changes in the intrinsic power
  of the central source, the bubble size and release frequency, at
  constant power, will vary with the surrounding gas
  properties. However the weakness of this dependence means this is
  likely to be an insignificant effect.

\item Models to date have taken no account of the possibility of the
  mixing of cool gas from the brightest cluster galaxy nebula into the
  ICM. We have shown that this process may plausibly have an effect on
  the temperature structure of the weak shock in Perseus, in which
  case it will alter the temperature structure produced by repeated
  heating episodes.

\item A non-homogeneous population of cosmic rays will affect how the
  energy dissipated by weak shock waves is distributed spatially.
\end{itemize}

In order to resolve these issues, it is clear that simulations with
repeated bubble formation events run over a period of several Gyr in
multiple dimensions are required. \citet{DallaVecchia2004} have run a
3D simulation for $1.5\gyr$ in which they are able to balance heating
and cooling, although their results are dependent on the radius over
which their bubbles are distributed, with small injection radii
(consistent with observations) leading to a potentially problematic
increase in energy of the cluster core. However their simulation
contains neither bulk gas motion not the inflation of the bubbles
themselves. \citet{Sijacki2007} have also run a cosmological
simulation including mechanical feedback from AGN, with the energy and
radius of the bubbles coupled to the black hole accretion rate. They
find the injection is able to prevent catastrophic cooling in the
cluster core, whilst retaining a flat or decreasing temperature
profile in the centre.

\section{Energetics of the Shocked Region}

The amount of mechanical energy imparted to the cool core region by
the AGN is critical in determining whether the AGN alone can offset
the cooling flow to the observed levels in cluster cores.  If the
bubble is slowly inflated and thereafter expands adiabatically as it
rises buoyantly through the cluster core, it releases an energy of
approximately $(\gamma_\text{bubble}/(\gamma_\text{bubble}-1))pV$,
i.e.  its enthalpy \citep{Churazov2002}. Therefore studies of AGN
heating have typically assumed $E = 4pV_\text{observed}$, appropriate
if the gas inside the bubble is purely relativistic. With these
assumptions, authors have typically found that the available energy is
enough to offset the cooling in some but not all clusters
(e.g. \citealt{Birzan2004, Dunn2006, Rafferty2006, Birzan2006}). The
fraction of clusters for which heating is observed to balance cooling
at the present time is often used to infer a duty cycle for mechanical
energy input by the AGN.

Some simulations have suggested that the overall energy release is
substantially greater than assumed in these studies due to the
contribution from non-adiabatic processes
(e.g. \citealt{Binney2007}). The high quality data from the Perseus
allow us to see the high pressure rims around the inner bubbles,
allowing an observational estimate of the extra energy deposited by
irreversible processes during the bubble expansion.

Following \cite{Dunn2005} we assume that the Perseus bubbles are approximately
spherical, of radius $r_\text{inner}$ and that the shocked material is
confined to a spherical shell of radius $r_\text{outer}$ outside each
bubble. The energy dissipated by the shock is then:

\begin{equation}
E_\text{shock} = \frac{1}{\gamma_\text{gas}-1}
\iiint_\text{shell}(p_\text{shocked} - p_\text{unshocked})dV
\label{eqn:shell_energy}
\end{equation}

The value of $p_\text{unshocked}$ is, of course, an unknown quantity,
which we must extrapolate from the properties of the cluster outside
the shock. In order to minimise the uncertainty associated with this
extrapolation, we assume that both bubbles are identical and that the
pressure profile through the whole region surrounding the bubble, both
inside and outside the shock, is well approximated by the profile in
that region in the sector we have studied in this paper. With these
assumptions we only need extrapolate the pressure profile over the
width of the high pressure region. To perform this extrapolation, we
use a power-law fit to the pressure profile in the range $20-45\kpc$,
measured from the bubble centre:

\begin{equation}
p_\text{unshocked} = 0.40\left(\frac{r}{1\kpc}\right)^{-0.16} \kev \cm^{-3},
\end{equation}

For the shocked pressure we used the values measured in the high
pressure region $r=10.1 - 19.3\kpc$ directly (where $r$ is measured
from the centre of the bubble as in Fig. \ref{fig:deprojT}). With
these parameters, the excess energy in the high pressure shell is
$1.6\times10^{59}\erg$ per bubble. If we estimate the power provided
by the shock as $E_\text{shock}/t_\text{shock}$ where
$t_\text{shock}=(r_\text{outer} - r_\text{inner})/Mc$, we find
$P_\text{shock}\sim6\times10^{44}\erg \s^{-1}$ per bubble. 

From \citet{Dunn2004}, the $pV$ work done in expanding the bubbles
adiabatically are $3.9\times10^{58}\erg$ and $5.3\times10^{58}\erg$
for the northern and southern bubbles, respectively. Therefore,
averaging over the two bubbles, we find that the energy in the shocked
region is roughly 3.5 times the $pV$ work done in expanding the
bubbles adiabatically, already comparable to the enthaphy of the
bubbles. Assuming this result generalises to other clusters, estimates
of the total AGN-supplied mechanical power available to heat cluster
cores which assume $4pV$ energy per bubble over their lifetime
are likely to be reliable.

The calculated power is somewhat higher than the range
$6\times10^{43}-2\times10^{44} \erg \s^{-1}$ per bubble found by
\citet{Dunn2004} on the basis of the work done expanding the bubbles
(i.e. taking the energy to be $pV$, not $4pV$) and various estimates
of the timescale for energy dissipation. It is also comparable to,
although, again, slightly larger than, the power in the ripples calculated by
\cite{Sanders2007}.

These estimates have all been made on the assumption that only the thermal
pressure is significant. \citet{Sanders2007} found that inside $40\kpc$
Perseus has a significant non-thermal pressure component contributing
about 50 percent of the thermal pressure. Therefore the energy inside the
shocked region may be an additional factor $\sim1.5$ higher than derived here.


It is also possible that excess energy has been channelled into other
modes such as gravitational energy, kinetic energy or turbulence. We
expect the contribution of gravitational energy to be small;
assuming the expansion is approximately symmetric, the almost linear
nature of the cluster potential over the region $r < r_\text{outer}$
means that the net change in the gravitational potential
cancels. Relaxing the condition of symmetrical redistribution, the
condition of hydrostatic equilibrium gives $\Delta P \sim \rho g \Delta
r$ so $E_\text{grav}/E_\text{thermal}\sim\Delta P/P \sim 0.1$
therefore the gravitational contribution to the total energy is likely
to be small. 

\section{Discussion and Conclusions}

We have performed a detailed analysis of the weak shock feature seen
in the core of the Perseus cluster. Deprojection of a sector
containing the shock shows no temperature rise coincident with the
density jump. There is a small rise $3\kpc$ downstream from the
shock front.  Models of shock production in spherically-symmetric,
ideal gas, cluster atmospheres produce poor fits to the observed
profile, principally because the observed postshock temperature is
lowest at the shock front, just where it is expected to be highest.

The presence of thermal conduction, or a reduced value of the
adiabatic index from the presence of a population of cosmic rays, can
produce a temperature jump with very similar magnitude to that seen in
the data and substantially improve the fit of the shock profile to the
data. However we did not find a set of parameters for which the
$\chi^2$ value of the fit is statistically acceptable. The difficulty
of finding a simple solution to this problem is likely related to the
fact that the densities and length scales relevant to the Perseus
shock represent the border between the collisional and collisionless
regimes. Therefore the non-hydrodynamic physics, such as the
interaction of the electron population with the local magnetic field,
may be essential to the formation of the observed structures.

Mixing of cold gas may have a significant effect on the energy
distribution in all cool core clusters. Of the nearby large cooling
flows, it is only Abell 2029 which does not show evidence for a large
filamentary nebula \citep{Johnstone1987}. In the Virgo cluster, where
both H$\alpha$ filaments \citep{Sparks2004} and a weak shock
\citep{Forman2006} are seen, the shock is outside the region of strong
filamentary emission. Unlike Perseus, there is no evidence of
deviation from the standard Rankine-Hugoniot relations for the shock
in Virgo.

Finally, we must consider the possibility that the pressure across the
density jump is in fact continuous with an increase in the thermal gas
pressure fraction in the ``preshock'' region. This could be the case
if the medium outside the front has a larger component of ram,
magnetic field or cosmic ray pressure, for example.

The geometry of the density jump seems to disfavour an increase in ram pressure;
because the shock is a continuous feature around the inner region
it is difficult to maintain a bulk flow of the material in the whole
outer region. 

The viability of the observed feature being produced by a change in
the magnetic field strength across the density front depends greatly
on the field strength in the inner region. \citet{Taylor2006} find a
magnetic field of around $25\mug$ based on rotation measure
measurements of the very inner regions of the cluster core, although
they suggest this field is associated largely with the filamentary
nebula, whilst \citet{Sanders2005} estimate a magnetic field of around
$0.1\mug$ from non-thermal X-ray emission. In order to provide a
continuous pressure across the shock, we require $B_\text{outer}^2 -
B_\text{inner}^2 \sim 6\times10^3 \mug^2$. If the inner field is
around $25\mug$ this is a factor of around $3$ increase in the field
strength. For a smaller outer field the required increase is
correspondingly larger and for a $0.1\mug$ inner field, magnetic
pressure balance across the density front would require an implausible
increase in the field strength of a factor of $\sim 1000$.

If there were a significant change in either the magnetic field or
non-thermal particle population in the gas moving across the density
front, we would expect to see a corresponding feature in the radio
mini-halo emission. No-such feature can obviously be identified in the
VLA data of \citet{Sijbring1993}, although we cannot entirely rule out
a model in which the non-thermal gas is the largest fraction by volume
and the ``shock'' feature is associated with a change in the number
density of clouds of thermal gas. Nevertheless, the difficulties
associated with these constant pressure models lead us to believe that
an outward-propagating weak shock wave is the most likely explanation
for the observed feature.

In conclusion, the high pressure regions surrounding the inner radio
bubbles in the Perseus cluster terminate in spherical shocks. The
density jumps by 31 percent at the shock front, yet any immediate
temperature jump is only $3\pm6$ percent. This is inconsistent with a
simple adiabatic shock. We are unable to find a definitive solution to
the problem. Several possibilities have been investigated, with
turbulent post-shock gas rapidly mixing with cool material associated
with the optical filaments being promising. Such mixing, and the 3D
geometry of the bubbles, can prevent the central gas from overheating,
despite repeated bubbling over hundreds of millions of years. The
excess thermal energy contained in these high pressure region is about
3.5 times the total work done in expanding the bubbles adiabatically
and the power provided by the shock is enough to significantly offset
the cooling in Perseus.

\section{Acknowledgements}

JG and ACF acknowledge support from the Science and Technology
Facilities Council and the Royal Society, respectively.

\bibliographystyle{mnras} 
\bibliography{refs}

\begin{thebibliography}{}

\bibitem[\protect\citeauthoryear{{Binney}, {Bibi}, \& {Omma}}{{Binney}
  et~al.}{2007}]{Binney2007}
{Binney} J., {Bibi} F.~A.,  {Omma} H., 2007, \mnras, 377, 142

\bibitem[\protect\citeauthoryear{{B{\^i}rzan} et~al.}{{B{\^i}rzan}
  et~al.}{2006}]{Birzan2006}
{B{\^i}rzan} L., {McNamara} B.~R., {Carilli} C.~L., {Nulsen} P.~E.~J.,  {Wise}
  M.~W., 2006, astro-ph/0612393

\bibitem[\protect\citeauthoryear{{B{\^i}rzan} et~al.}{{B{\^i}rzan}
  et~al.}{2004}]{Birzan2004}
{B{\^i}rzan} L., {Rafferty} D.~A., {McNamara} B.~R., {Wise} M.~W.,  {Nulsen}
  P.~E.~J., 2004, \apj, 607, 800

\bibitem[\protect\citeauthoryear{{Borkowski}, {Lyerly}, \&
  {Reynolds}}{{Borkowski} et~al.}{2001}]{Borkowski2001}
{Borkowski} K.~J., {Lyerly} W.~J.,  {Reynolds} S.~P., 2001, \apj, 548, 820

\bibitem[\protect\citeauthoryear{{Borkowski}, {Sarazin}, \&
  {Blondin}}{{Borkowski} et~al.}{1994}]{Borkowski1994}
{Borkowski} K.~J., {Sarazin} C.~L.,  {Blondin} J.~M., 1994, \apj, 429, 710

\bibitem[\protect\citeauthoryear{{Br{\"u}ggen}, {Ruszkowski}, \&
  {Hallman}}{{Br{\"u}ggen} et~al.}{2005}]{Bruggen2005}
{Br{\"u}ggen} M., {Ruszkowski} M.,  {Hallman} E., 2005, \apj, 630, 740

\bibitem[\protect\citeauthoryear{{Churazov} et~al.}{{Churazov}
  et~al.}{2002}]{Churazov2002}
{Churazov} E., {Sunyaev} R., {Forman} W.,  {B{\"o}hringer} H., 2002, \mnras,
  332, 729

\bibitem[\protect\citeauthoryear{{Conselice}, {Gallagher}, \&
  {Wyse}}{{Conselice} et~al.}{2001}]{Conselice2001}
{Conselice} C.~J., {Gallagher} J.~S.,  {Wyse} R.~F.~G., 2001, \aj, 122, 2281

\bibitem[\protect\citeauthoryear{{Cowie} \& {McKee}}{{Cowie} \&
  {McKee}}{1977}]{CowieMckee1977}
{Cowie} L.~L.,  {McKee} C.~F., 1977, \apj, 211, 135

\bibitem[\protect\citeauthoryear{{Dalla Vecchia} et~al.}{{Dalla Vecchia}
  et~al.}{2004}]{DallaVecchia2004}
{Dalla Vecchia} C., {Bower} R.~G., {Theuns} T., {Balogh} M.~L., {Mazzotta} P.,
  {Frenk} C.~S., 2004, \mnras, 355, 995

\bibitem[\protect\citeauthoryear{{Dunn} \& {Fabian}}{{Dunn} \&
  {Fabian}}{2004}]{Dunn2004}
{Dunn} R.~J.~H.,  {Fabian} A.~C., 2004, \mnras, 355, 862

\bibitem[\protect\citeauthoryear{{Dunn} \& {Fabian}}{{Dunn} \&
  {Fabian}}{2006}]{Dunn2006}
{Dunn} R.~J.~H.,  {Fabian} A.~C., 2006, \mnras, 373, 959

\bibitem[\protect\citeauthoryear{{Dunn}, {Fabian}, \& {Taylor}}{{Dunn}
  et~al.}{2005}]{Dunn2005}
{Dunn} R.~J.~H., {Fabian} A.~C.,  {Taylor} G.~B., 2005, \mnras, 364, 1343

\bibitem[\protect\citeauthoryear{{Fabian}}{{Fabian}}{1994}]{Fabian1994}
{Fabian} A.~C., 1994, \araa, 32, 277

\bibitem[\protect\citeauthoryear{{Fabian} et~al.}{{Fabian}
  et~al.}{2003a}]{Fabian2003}
{Fabian} A.~C., {Sanders} J.~S., {Allen} S.~W., {Crawford} C.~S., {Iwasawa} K.,
  {Johnstone} R.~M., {Schmidt} R.~W.,  {Taylor} G.~B., 2003a, \mnras, 344, L43

\bibitem[\protect\citeauthoryear{{Fabian} et~al.}{{Fabian}
  et~al.}{2003b}]{FabianHalpha}
{Fabian} A.~C., {Sanders} J.~S., {Crawford} C.~S., {Conselice} C.~J.,
  {Gallagher} J.~S.,  {Wyse} R.~F.~G., 2003b, \mnras, 344, L48

\bibitem[\protect\citeauthoryear{{Fabian} et~al.}{{Fabian}
  et~al.}{2006}]{Fabian2006}
{Fabian} A.~C., {Sanders} J.~S., {Taylor} G.~B., {Allen} S.~W., {Crawford}
  C.~S., {Johnstone} R.~M.,  {Iwasawa} K., 2006, \mnras, 366, 417

\bibitem[\protect\citeauthoryear{{Forman} et~al.}{{Forman}
  et~al.}{2007}]{Forman2006}
{Forman} W. et~al., 2007, \apj, 665, 1057

\bibitem[\protect\citeauthoryear{{Fujita} \& {Suzuki}}{{Fujita} \&
  {Suzuki}}{2006}]{Fujita2006}
{Fujita} Y.,  {Suzuki} T.~K., 2006, ArXiv Astrophysics e-prints

\bibitem[\protect\citeauthoryear{{Gardini}}{{Gardini}}{2007}]{Gardini2007}
{Gardini} A., 2007, \aap, 464, 143

\bibitem[\protect\citeauthoryear{{Hamilton}, {Chevalier}, \&
  {Sarazin}}{{Hamilton} et~al.}{1983}]{Hamilton1983}
{Hamilton} A.~J.~S., {Chevalier} R.~A.,  {Sarazin} C.~L., 1983, \apjs, 51, 115

\bibitem[\protect\citeauthoryear{{Hatch} et~al.}{{Hatch}
  et~al.}{2005}]{Hatch2005}
{Hatch} N.~A., {Crawford} C.~S., {Fabian} A.~C.,  {Johnstone} R.~M., 2005,
  \mnras, 358, 765

\bibitem[\protect\citeauthoryear{{Heinz} et~al.}{{Heinz}
  et~al.}{2006}]{Heinz2006}
{Heinz} S., {Br{\"u}ggen} M., {Young} A.,  {Levesque} E., 2006, \mnras, 373,
  L65

\bibitem[\protect\citeauthoryear{{Johnstone}, {Fabian}, \&
  {Nulsen}}{{Johnstone} et~al.}{1987}]{Johnstone1987}
{Johnstone} R.~M., {Fabian} A.~C.,  {Nulsen} P.~E.~J., 1987, \mnras, 224, 75

\bibitem[\protect\citeauthoryear{{Johnstone} et~al.}{{Johnstone}
  et~al.}{2007}]{Johnstone2007}
{Johnstone} R.~M., {Hatch} N.~A., {Ferland} G.~J., {Fabian} A.~C., {Crawford}
  C.~S.,  {Wilman} R.~J., 2007, \mnras, 1005

\bibitem[\protect\citeauthoryear{{Landau} \& {Lifshitz}}{{Landau} \&
  {Lifshitz}}{1959}]{LandauLifshitzFluids}
{Landau} L.~D.,  {Lifshitz} E.~M., 1959, {Fluid mechanics}.
\newblock Course of theoretical physics, Oxford: Pergamon Press, 1959

\bibitem[\protect\citeauthoryear{{Liedahl}, {Osterheld}, \&
  {Goldstein}}{{Liedahl} et~al.}{1995}]{Liedahl1995}
{Liedahl} D.~A., {Osterheld} A.~L.,  {Goldstein} W.~H., 1995, \apjl, 438, L115

\bibitem[\protect\citeauthoryear{{Mathews} \& {Brighenti}}{{Mathews} \&
  {Brighenti}}{2007}]{Mathews2007}
{Mathews} W.~G.,  {Brighenti} F., 2007, \apj, 660, 1137

\bibitem[\protect\citeauthoryear{{Mathews}, {Faltenbacher}, \&
  {Brighenti}}{{Mathews} et~al.}{2006}]{Mathews2006}
{Mathews} W.~G., {Faltenbacher} A.,  {Brighenti} F., 2006, \apj, 638, 659

\bibitem[\protect\citeauthoryear{{Morris}}{{Morris}}{2003}]{Morris2003}
{Morris} R., 2003, Ph.D. thesis, University of Cambridge

\bibitem[\protect\citeauthoryear{{Narayan} \& {Medvedev}}{{Narayan} \&
  {Medvedev}}{2001}]{Narayan2001}
{Narayan} R.,  {Medvedev} M.~V., 2001, \apjl, 562, L129

\bibitem[\protect\citeauthoryear{{Navarro}, {Frenk}, \& {White}}{{Navarro}
  et~al.}{1997}]{NFW1997}
{Navarro} J.~F., {Frenk} C.~S.,  {White} S.~D.~M., 1997, \apj, 490, 493

\bibitem[\protect\citeauthoryear{{Nipoti} \& {Binney}}{{Nipoti} \&
  {Binney}}{2004}]{NipotiBinney2004}
{Nipoti} C.,  {Binney} J., 2004, \mnras, 349, 1509

\bibitem[\protect\citeauthoryear{{Peterson} et~al.}{{Peterson}
  et~al.}{2003}]{Peterson2003}
{Peterson} J.~R., {Kahn} S.~M., {Paerels} F.~B.~S., {Kaastra} J.~S., {Tamura}
  T., {Bleeker} J.~A.~M., {Ferrigno} C.,  {Jernigan} J.~G., 2003, \apj, 590,
  207

\bibitem[\protect\citeauthoryear{{Peterson} et~al.}{{Peterson}
  et~al.}{2001}]{Peterson2001}
{Peterson} J.~R. et~al., 2001, \aap, 365, L104

\bibitem[\protect\citeauthoryear{{Pfrommer} et~al.}{{Pfrommer}
  et~al.}{2007}]{Pfrommer2007}
{Pfrommer} C., {En{\ss}lin} T.~A., {Springel} V., {Jubelgas} M.,  {Dolag} K.,
  2007, \mnras, 430

\bibitem[\protect\citeauthoryear{{Rafferty} et~al.}{{Rafferty}
  et~al.}{2006}]{Rafferty2006}
{Rafferty} D.~A., {McNamara} B.~R., {Nulsen} P.~E.~J.,  {Wise} M.~W., 2006,
  \apj, 652, 216

\bibitem[\protect\citeauthoryear{{Ritchmyer} \& {Morton}}{{Ritchmyer} \&
  {Morton}}{1967}]{RichtmyerMorton}
{Ritchmyer} R.~D.,  {Morton} K.~W., 1967, {Difference methods for initial-value
  problems}.
\newblock Interscience Tracts in Pure and Applied Mathematics, New York:
  Interscience, 1967, 2nd ed.

\bibitem[\protect\citeauthoryear{{Ruszkowski} et~al.}{{Ruszkowski}
  et~al.}{2007a}]{Ruszkowski2007a}
{Ruszkowski} M., {Ensslin} T.~A., {Bruggen} M., {Begelman} M.~C.,  {Churazov}
  E., 2007a, ArXiv e-prints, 705

\bibitem[\protect\citeauthoryear{{Ruszkowski} et~al.}{{Ruszkowski}
  et~al.}{2007b}]{Ruszkowski2007}
{Ruszkowski} M., {En{\ss}lin} T.~A., {Br{\"u}ggen} M., {Heinz} S.,  {Pfrommer}
  C., 2007b, \mnras, 378, 662

\bibitem[\protect\citeauthoryear{{Salom{\'e}} et~al.}{{Salom{\'e}}
  et~al.}{2006}]{Salome2006}
{Salom{\'e}} P. et~al., 2006, \aap, 454, 437

\bibitem[\protect\citeauthoryear{{Sanders} \& {Fabian}}{{Sanders} \&
  {Fabian}}{2007}]{Sanders2007}
{Sanders} J.~S.,  {Fabian} A.~C., 2007, \mnras, 381, 1381

\bibitem[\protect\citeauthoryear{{Sanders} et~al.}{{Sanders}
  et~al.}{2004}]{Sanders2004}
{Sanders} J.~S., {Fabian} A.~C., {Allen} S.~W.,  {Schmidt} R.~W., 2004, \mnras,
  349, 952

\bibitem[\protect\citeauthoryear{{Sanders}, {Fabian}, \& {Dunn}}{{Sanders}
  et~al.}{2005}]{Sanders2005}
{Sanders} J.~S., {Fabian} A.~C.,  {Dunn} R.~J.~H., 2005, \mnras, 360, 133

\bibitem[\protect\citeauthoryear{{Sijacki} et~al.}{{Sijacki}
  et~al.}{2007}]{Sijacki2007}
{Sijacki} D., {Springel} V., {di Matteo} T.,  {Hernquist} L., 2007, \mnras,
  380, 877

\bibitem[\protect\citeauthoryear{{Sijbring}}{{Sijbring}}{1993}]{Sijbring1993}
{Sijbring} D., 1993, Ph.D. thesis, Groningen

\bibitem[\protect\citeauthoryear{{Soker} \& {Sarazin}}{{Soker} \&
  {Sarazin}}{1990}]{SokerSarazin1990}
{Soker} N.,  {Sarazin} C.~L., 1990, \apj, 348, 73

\bibitem[\protect\citeauthoryear{{Sparks} et~al.}{{Sparks}
  et~al.}{2004}]{Sparks2004}
{Sparks} W.~B., {Donahue} M., {Jord{\'a}n} A., {Ferrarese} L.,  {C{\^o}t{\'e}}
  P., 2004, \apj, 607, 294

\bibitem[\protect\citeauthoryear{{Spitzer}}{{Spitzer}}{1956}]{SpitzerFullyIoni%
zed}
{Spitzer} L., 1956, {Physics of Fully Ionized Gases}.
\newblock Physics of Fully Ionized Gases, New York: Interscience Publishers,
  1956

\bibitem[\protect\citeauthoryear{{Taylor} et~al.}{{Taylor}
  et~al.}{2006}]{Taylor2006}
{Taylor} G.~B., {Gugliucci} N.~E., {Fabian} A.~C., {Sanders} J.~S., {Gentile}
  G.,  {Allen} S.~W., 2006, \mnras, 368, 1500

\bibitem[\protect\citeauthoryear{{Zel'Dovich} \& {Raizer}}{{Zel'Dovich} \&
  {Raizer}}{1967}]{ZeldovichRaizer}
{Zel'Dovich} Y.~B.,  {Raizer} Y.~P., 1967, {Physics of shock waves and
  high-temperature hydrodynamic phenomena}.
\newblock New York: Academic Press, 1966/1967, edited by Hayes, W.D.;
  Probstein, Ronald F.

\end{thebibliography}

\end{document}